\documentclass[a4paper,12pt]{article}
\usepackage{amssymb}
\usepackage{amsmath,amsfonts,epsfig,subfigure}
\numberwithin{equation}{section}
\begin{document}
\title{Surface waves in a stretched and sheared \\
incompressible elastic material}

\author{Michel Destrade, Ray W. Ogden}
\date{2005}
\maketitle

\bigskip

%

\begin{abstract}
\noindent In this paper we analyze the effect of a combined pure
homogeneous strain and simple shear in a principal plane of the
latter on the propagation of surface waves for an incompressible
isotropic elastic half-space whose boundary is normal to the glide
planes of the shear.  This generalizes previous work in which,
separately, pure homogeneous strain and simple shear were
considered.  For a special class of materials the secular equation
is obtained in explicit form and then specialized to recover
results obtained previously for the two cases mentioned above. A
method for obtaining the secular equation for a general form of
strain-energy function is then outlined.  In general this is very
lengthy and the result is not listed, but, for the case in which
there is no normal stress on the half-space boundary, the result
is given, for illustration, in respect of the so-called
generalized Varga material. Numerical results are given to show
how the surface wave speed depends on both the underlying pure
homogeneous strain and the superimposed simple shear. Further
numerical results are provided for the Gent model of limiting
chain extensibility.

\end{abstract}

\newpage 

\section{Introduction}

The vibration and wave propagation characteristics of rubberlike
solids are important for practical applications in, for example,
engine mountings and seismic isolators, and an understanding of
these characteristics is therefore of crucial importance.  There
have been many contributions in recent years to the analysis of,
in particular, vibrations and surface and interfacial waves in
finitely deformed elastic bodies, but relatively little attention
has been given to shearing deformations, for which the orientation
of the principal axes of deformation depends on the magnitude of
the shear.  In this paper we are concerned with the influence of a
simple shear combined with a pure homogeneous strain on the
propagation of surface waves.

Analysis of surface waves in a half-space subject to pure
homogeneous deformation dates back to the classic paper of Hayes
and Rivlin \cite{Hayes61}, which was concerned primarily with
compressible materials, while Flavin \cite{Flav63} considered the
same problem for an incompressible material, specifically the
Mooney-Rivlin material (for completeness, mention should be made
of early and noteworthy articles by Biot \cite{Biot40} and by
Buckens \cite{Buck58} and the related work of Biot summarized in
\cite{Biot65}). A more detailed analysis was provided for general
incompressible elastic materials by Dowaikh and Ogden
\cite{DoOg90}, who also examined the connection between surface
wave propagation and stability of the half-space, again for the
case of pure homogeneous strain.  We refer to this latter paper,
that by Chadwick \cite{Chad97} and the recent contribution by Fu
\cite{Fu04}, for pointers to the relevant literature. The only
paper thus far that deals with the influence of shear on surface
wave propagation appears to be that by Connor and Ogden
\cite{CoOg95}, and the closely related paper \cite{CoOg96}. These
are concerned with simple shear and give explicit results for a
certain class of incompressible materials, of which the best known
representative is the Mooney-Rivlin model for rubberlike solids.

The purpose of this paper is to extend the analysis in the above
papers to the situation in which a simple shear is combined with a
pure homogeneous deformation and for a general incompressible
isotropic hyperelastic material. These extensions are motivated by
several factors. First, one can easily picture situations of
practical interest where a solid is stretched and sheared: for
instance, a rubber isolator under a bridge is subjected to a
vertical compression and is then sheared as a result of thermal
extensions and contractions of the roadway. Also, the combination
of stretch and shear theoretically allows a square block of
incompressible material to be maintained in a deformed state with
no shear tractions nor normal tractions on its faces \cite{RaWi87}, in
contrast to simple shear and its associated Poynting effect.
Finally, the quite recent application of nonlinear elasticity to
the modelling of some biological tissues (see Humphrey
\cite{Hump02} or Holzapfel and Ogden \cite{HoOg03}) has
highlighted the need for solutions valid for more general
strain-energy functions because the functions used for
biomaterials are usually more complicated than those used for
rubberlike materials.

In Section 2 the basic equations for the (static) finite
deformation consisting of a pure homogeneous strain (or triaxial
stretch) followed by a simple shear in a principal plane of the
pure homogeneous strain are summarized for an incompressible
isotropic elastic material. The corresponding equations for
infinitesimal surface waves superimposed on this finite
deformation are then outlined. The equations and zero incremental
traction boundary conditions are applied to surface waves
propagating in the direction of shear and parallel to a half-space
whose normal lies within the plane of shear.  The secular equation
for the wave speed is then obtained in implicit form in respect of
a general strain-energy function.

This equation in examined in detail in Section 3.  For a class of
materials that includes the Mooney-Rivlin strain energy, the
secular equation is given in simple explicit form, which relates
the wave speed to the material parameters, the deformation and the
normal stress on the half-space associated with the underlying
finite deformation. Next, an alternative approach is outlined,
which enables the secular equation to be made explicit in the more
general case.  However, since this is very complicated we
illustrate the results by considering a generalized form of the
Varga material model.  Specifically, in the absence of normal
loads on the half-space boundary we obtain the secular equation in
relatively simple explicit form as a quartic for the squared wave
speed.  On the other hand, in the presence of normal loads we give
the corresponding explicit form of bifurcation condition
(corresponding to zero wave speed), which consists of two separate
equations. For each of the material models considered numerical
results are described in Section 4, illustrating the dependence of
the wave speed separately on the stretching and shearing parts of
the underlying deformation. Finally, for comparison, some limited
results are given for a model of finite chain extensibility of
rubberlike solids originally due to Gent \cite{Gent96}, and
recently used by Horgan and Saccomandi \cite{HoSa03} to model
strain-hardening arterial walls.

In closing this Introduction, we note that the method and the
results presented in the paper are easily extended to a generic
type of homogeneous pre-strain having a plane of symmetry, and to
unconstrained (compressible) solids. The assumptions of combined
stretch and shear and of incompressibility lead to some
simplifications and to neat algebraic expressions, but they are
inessential and should not be viewed as restrictive.

\section{Basic equations}
%
\subsection{Finite static deformations}
%
Let $(X_1,X_2,X_3)$ be a rectangular Cartesian coordinate system.
Consider a homogeneous semi-infinite body occupying the region
$X_2 \ge 0$ in its natural (unstressed) configuration. Suppose
that the body is composed of an incompressible isotropic elastic
material characterized by a constant mass density $\rho$ and a
strain-energy function $W(\lambda_1, \lambda_2, \lambda_3)$ (per
unit volume) written as a (symmetric) function of the (positive)
principal stretches $(\lambda_1,\lambda_2,\lambda_3)$, which
satisfy the incompressibility constraint
\begin{equation}\label{incomp}
\lambda_1 \lambda_2 \lambda_3=1.
\end{equation}

The body is subjected first to a finite pure homogeneous triaxial
stretch with principal stretch ratios $\mu_1, \mu_2, \mu_3$
($\mu_1 \mu_2 \mu_3 = 1$) and then to a finite simple shear of
amount $\kappa$, so that a particle initially at ($X_1, X_2, X_3$)
is displaced to ($\hat{x}_1, \hat{x}_2, \hat{x}_3$) according to
\begin{equation} \label{static}
\hat{x}_1 = \mu_1 X_1 + \kappa \mu_2 X_2, \quad
\hat{x}_2 = \mu_2 X_2, \quad
\hat{x}_3 = \mu_3 X_3.
\end{equation}

\begin{figure}
\centering \mbox{\subfigure[Unit cube]{\epsfig{figure=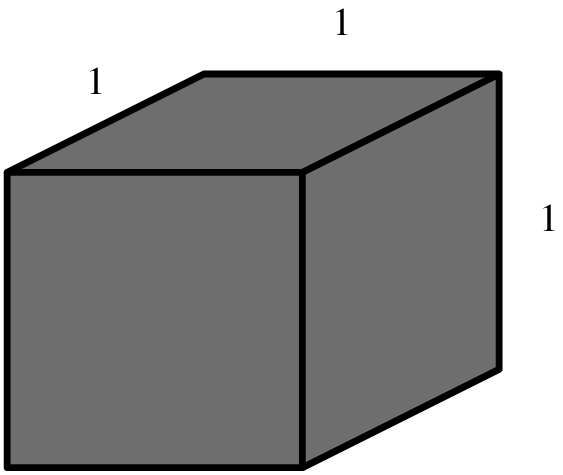,
width=.28\textwidth}}
  \quad \quad
     \subfigure[Triaxial stretch]{\epsfig{figure=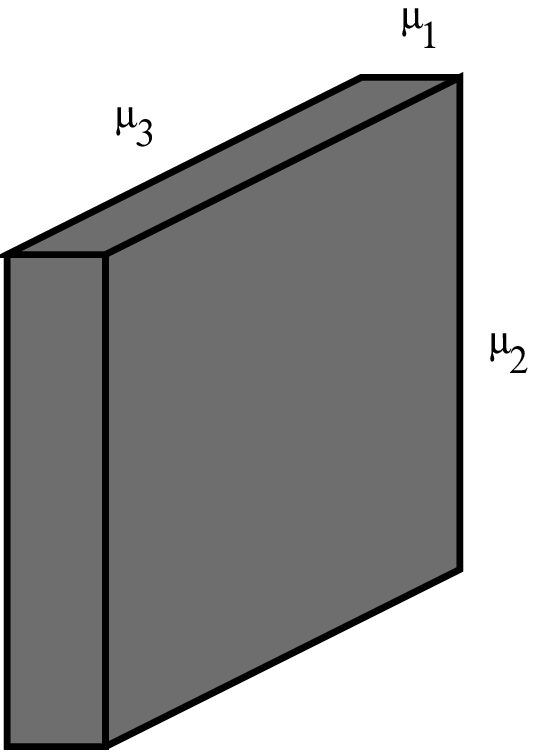,
width=.28\textwidth}}
  \quad \quad
     \subfigure[Simple shear]{\epsfig{figure=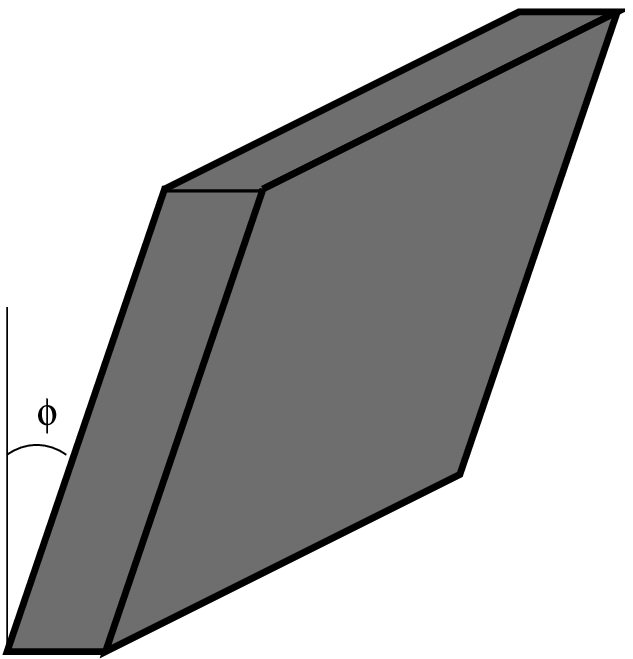,
width=.32\textwidth}}} \caption{Triaxial stretch followed by
simple shear of a unit cube.}
\end{figure}

Figure 1 shows the successive deformations of a unit cube in the
half-space; in Figure 1(c) the angle $\phi$ is the angle of shear,
which is given by $\tan \phi = \kappa$.

The deformation gradient $\mathbf{F}$ associated with the
deformation \eqref{static} has components
\begin{equation}\label{eff}
 \begin{bmatrix}
                         \mu_1 & \kappa \mu_2 & 0     \\
                           0   &  \mu_2       & 0     \\
                           0   &    0         & \mu_3
                        \end{bmatrix}.
\end{equation}
Let $x_1,x_2,x_3$, be the rectangular Cartesian coordinates
associated with the (orthogonal) principal axes of $\mathbf{FF}^T$
(the left Cauchy-Green deformation tensor), one of which is the
$\hat{x}_3$ direction, so that $x_3=\hat{x}_3$.  Within the
$(x_1,x_2)$ plane let the $x_1$ principal axis make an angle
$\theta$ with the $\hat{x}_1$ axis measured in the anticlockwise
sense.  Then the coordinates ($ \hat{x}_1, \hat{x}_2, \hat{x}_3$)
are related to ($x_1, x_2, x_3$) through a rotation according to
\begin{equation} \label{omega}
\hat{x}_i = \Omega_{ij} x_j, \quad
 [\Omega_{ij}] =
   \begin{bmatrix}
     \cos \theta & -\sin \theta & 0 \\
    \sin \theta & \cos \theta & 0 \\
          0      &      0      & 1
    \end{bmatrix}.
\end{equation}

The principal stretches, subject to \eqref{incomp}, and the angle $\theta$
are given in terms of the components \eqref{eff} by
\begin{align}
\label{theta} & \lambda_1^2 \cos^2 \theta + \lambda_2^2 \sin^2 \theta
 = \mu_1^2 + \mu_2^2 \kappa^2,
\nonumber \\
& (\lambda_1^2  - \lambda_2^2) \cos \theta \sin \theta
 = \mu_2^2 \kappa,
\nonumber \\
& \lambda_1^2 \sin^2 \theta + \lambda_2^2 \cos^2 \theta
 =  \mu_2^2,
\end{align}
from which we obtain
\begin{equation}
\lambda_1 \pm \lambda_2
 = \sqrt{(\mu_1 \pm \mu_2)^2 + \kappa^2\mu_2^2},
\quad
\lambda_3 = 1/(\mu_1 \mu_2),
\end{equation}
and
\begin{equation}
\tan 2 \theta
   = 2 \mu_2^2 \kappa / (\mu_1^2 - \mu_2^2 + \kappa^2\mu_2^2)
\end{equation}
(see, for example, \cite{WiGa84,RaWi87,BeHa92}). When the material
is sheared but not stretched ($\mu_1 = \mu_2 = \mu_3 = 1$), for
example, the well-known formulas
\begin{equation}
\lambda_{1,2} = (\sqrt{4 + \kappa^2} \pm \kappa)/2, \quad
\lambda_3 = 1, \quad \tan 2 \theta =  2 / \kappa,
\end{equation}
are recovered.

Let $\sigma_1,\sigma_2,\sigma_3$ denote the principal Cauchy
stresses.  Since the material is isotropic the principal axes of
Cauchy stress coincide with those of $\mathbf{FF}^T$, and hence,
analogously to \eqref{theta}, we may express the components of the
Cauchy stress on the axes associated with the coordinates ($
\hat{x}_1, \hat{x}_2, \hat{x}_3$), denoted by $\hat{\sigma}_{ij}$,
as
\begin{align} \label{sigma}
& \hat{\sigma}_{11}=\sigma_1 \cos^2 \theta + \sigma_2 \sin^2
\theta,
\nonumber \\
& \hat{\sigma}_{12}=(\sigma_1  - \sigma_2) \sin \theta \cos
\theta,
\nonumber \\
& \hat{\sigma}_{22}=\sigma_1 \sin^2 \theta + \sigma_2 \cos^2
\theta,
\end{align}
with
$\hat{\sigma}_{13}=\hat{\sigma}_{23}=0,\,\hat{\sigma}_{33}=\sigma_3$.

In terms of the strain-energy function, the principal Cauchy
stresses are given by
\begin{equation}
\sigma_i=\lambda_i\frac{\partial W}{\partial \lambda_i}-p,\quad
i\in\{1,2,3\}\quad \mbox{no sum over $i$},
\end{equation}
where $p$ is a Lagrange multiplier associated with the constrain
\eqref{incomp}.

The initial half-space $X_2\geq 0$ becomes the half-space
$\hat{x}_2\geq 0$ after the deformation described above, and the
load on its boundary $\hat{x}_2= 0$ has normal component
$\hat{\sigma}_{22}$ and shear component $\hat{\sigma}_{12}$.  We
now consider the propagation of surface waves in this half-space.

%
\subsection{Superimposed infinitesimal surface waves}
%
Now consider a surface (Rayleigh) wave propagating with speed $v$
and wave number $k$ in the $\hat{x}_1$ direction in the considered
deformed half-space ($\hat{x}_2\geq 0$), with attenuation in the
$\hat{x}_2$ direction. We emphasize, as noted above and in
contrast to many previous studies of surface waves in deformed
materials, these directions do not coincide with the principal
directions $(x_1,x_2)$ of strain for the underlying
pre-deformation. It seems that the only paper considering the
propagation of non-principal surface waves over a non-principal
plane is that by Connor and Ogden \cite{CoOg95}, where the problem
was solved for a sheared half-space ($\mu_1=\mu_2=\mu_3=1$) and
for a specific class of materials (see also the related paper by
Connor and Ogden \cite{CoOg96}).

Let the incremental displacement $\mathbf{u}$ associated with the
wave have components $\hat{u}_i,\, i\in\{1,2,3\}$ in the
coordinates system ($\hat{x}_1, \hat{x}_2, \hat{x}_3$), with
dependence, in general, on $(\hat{x}_1,\hat{x}_2,\hat{x}_3,t)$,
where $t$ is time. The corresponding components $\hat{s}_{ij}$ of
the incremental nominal stress tensor are then given by
\begin{equation} \label{s}
\hat{s}_{ij} =
 \hat{B}_{ijkl} \frac{\partial \hat{u}_l}{\partial \hat{x}_k}
 +p\frac{\partial \hat{u}_i}{\partial \hat{x}_j}
   - \hat{p} \delta_{ij},
\end{equation}
where $\hat{B}_{ijkl}$ are the components of $\mathbf{B}$, the
tensor of first-order elastic moduli associated with the
underlying deformation, and $\hat{p}$ is the incremental
counterpart of $p$. The non-vanishing components of the elasticity
tensor $\mathbf{B}$, denoted $B_{ijkl}$, in the coordinate system
($x_1, x_2, x_3$) are given by
\begin{align}
& B_{iijj} = \lambda_i \lambda_j W_{ij} ,
\nonumber \\
& B_{ijij} =
  (\lambda_i W_i-\lambda_j
  W_j)\lambda_i^2/(\lambda_i^2-\lambda_j^2)\quad i\neq j,
\nonumber \\
& B_{ijji}=  B_{jiij} =
   B_{ijij} - \lambda_i W_i \quad i\neq j
&&
\end{align}
(see, for example, \cite{Ogde84,DoOg90,ChWB85}), where $W_i :=
\partial W/\partial\lambda_i,\, W_{ij} :=
\partial^2 W/\partial\lambda_i\partial\lambda_j$. Note that
expressions for $B_{ijij}$ in the case $\lambda_i=\lambda_j,\,
i\neq j$, may be obtained by a limiting process.

The two sets of components of $\mathbf{B}$ are related by
\begin{equation} \label{rotation}
\hat{B}_{ijkl} = \Omega_{ip}\Omega_{jq}\Omega_{kr}\Omega_{ls}
B_{pqrs}
\end{equation}
and they satisfy the symmetry properties
\begin{equation}
B_{ijkl} = B_{klij}, \quad \hat{B}_{ijkl} = \hat{B}_{klij}.
\end{equation}

Also, note that for $i,j\in\{ 1,2\}$, in particular, $\mathbf{B}$
has six non-zero entries in the $(x_1,x_2,x_3)$ system, namely
$B_{1111}$, $B_{2222}$, $B_{1122}$, $B_{1221}$, $B_{1212}$, and
$B_{2121}$, whereas in the $(\hat{x}_1,\hat{x}_2,\hat{x}_3)$
system it has ten: $\hat{B}_{1111}$, $\hat{B}_{2222}$,
$\hat{B}_{1122}$, $\hat{B}_{1221}$, $\hat{B}_{1212}$,
$\hat{B}_{2121}$, the counterparts of those above, together with
$\hat{B}_{1112}$, $\hat{B}_{1121}$, $\hat{B}_{1222}$, and
$\hat{B}_{2122}$.

For surface waves propagating on a half-space subject to a pure
homogeneous stretch and no shear ($\kappa =0$), Dowaikh and Ogden
\cite{DoOg90} found it convenient to introduce the shorthand
notations
\begin{equation}
\alpha = B_{1212}, \quad
\gamma = B_{2121} = \lambda_1^{-2}\lambda_2^2 \alpha, \quad
2\beta = B_{1111} + B_{2222} - 2B_{1122} - 2B_{1221}.
\end{equation}
The corresponding quantities in the
$(\hat{x}_1,\hat{x}_2,\hat{x}_3)$ system are obtained by using
\eqref{rotation} in the form
\begin{align} \label{shortHand}
\hat{\alpha}  & := \hat{B}_{1212}
      = \alpha \cos^4 \theta
            + 2\beta \sin^2 \theta \cos^2 \theta
                + \gamma \sin^4 \theta,
\nonumber \\
 \hat{\gamma} & := \hat{B}_{2121}
  = \alpha \sin^4 \theta
       + 2\beta  \sin^2 \theta \cos^2 \theta
        + \gamma \cos^4 \theta,
\nonumber \\
 2\hat{\beta} & := \hat{B}_{1111} +\hat{B}_{2222}
                              -2\hat{B}_{1122} -2\hat{B}_{1221}
= 2 \beta
           - 6(2\beta - \alpha - \gamma) \sin^2 \theta \cos^2 \theta.
\end{align}
It is easily checked that neither $\alpha$, $\gamma$, $\beta$ nor
$\hat{\alpha}$, $\hat{\gamma}$, $\hat{\beta}$ depend separately on
the constant Cauchy stress components
($\sigma_1,\sigma_2,\sigma_3$) (or equivalently on
$\hat{\sigma}_{ij}$). It follows from \eqref{shortHand} that
\begin{equation}
\hat{\alpha}-\hat{\gamma}=(\alpha-\gamma)\cos 2\theta,\quad
\hat{\alpha} + \hat{\gamma} - 2 \hat{\beta}=(\alpha + \gamma -
2\beta)\cos 4\theta.
\end{equation}
From the latter it follows that if the equality $\alpha + \gamma -
2\beta = 0$ is satisfied, as it is for certain forms of the
strain-energy function, then we also have $\hat{\alpha} +
\hat{\gamma} - 2 \hat{\beta} = 0$ for any angle $\theta$. In such
a case the equality is said to be `structurally invariant'
\cite{Ting00}.

Finally in this section, we note that the \textit{strong
ellipticity} condition may be expressed in the form
\begin{equation} \label{SE}
\alpha >0, \quad \beta + \sqrt{\alpha \gamma} >0
\end{equation}
(see, for example, Dowaikh and Ogden \cite{DoOg90} and Chadwick
\cite{Chad97}). It follows immediately that $\hat{\alpha}
>0$, but the counterpart of the second inequality in \eqref{SE}
for the $(\hat{x}_1,\hat{x}_2,\hat{x}_3)$ system is not so simple,
and is not therefore given here.

\subsection{Equations of motion and boundary conditions}

We now restrict attention to waves of the form
\begin{equation} \label{wave}
\mathbf{u} (\hat{x}_1, \hat{x}_2, t)
  = \mathbf{U}(k\hat{x}_2) e^{{\rm i}k(\hat{x}_1 -vt)},
\end{equation}
independent of $\hat{x}_3$, where $\mathbf{U}$ is a function of
$k\hat{x}_2$, $k$ is the wave number and $v$ the wave speed. This
must satisfy the incremental equations of motion and the
incremental incompressibility constraint. Respectively, these are
\begin{equation} \label{motion}
\frac{\partial \hat{s}_{ji}}{\partial \hat{x}_j}
 = \rho \frac{\partial^2 \hat{u}_i}{\partial t^2},\quad
\hat{u}_{i,i} = 0 \quad \text{in}\ \hat{x}_2> 0.
\end{equation}
Additionally, we impose the condition of zero incremental traction
on the plane boundary, so that
\begin{equation} \label{BC}
\hat{s}_{2i} = 0\quad \text{on}\ \hat{x}_2 = 0, \quad i=1,2,3.
\end{equation}

In view of equations \eqref{wave} and \eqref{s}, $\hat{s}_{ij}$
and $\hat{p}$ take forms similar to that of the displacement
$\mathbf{u}$, that is
\begin{equation} \label{fields}
\{\hat{s}_{ij}(\hat{x}_1,\hat{x}_2,t) ,
\hat{p}(\hat{x}_1,\hat{x}_2,t)\} = \{
        kS_{ij}(k\hat{x}_2), {\rm i}k P(k\hat{x}_2) \}
                                          e^{{\rm i}k(\hat{x}_1 -
                                          vt)},
\end{equation}
where we have introduced the functions $S_{ij}$ and $P$ of
$k\hat{x}_2$ It is a straightforward exercise to check that the
in-plane equations decouple from the out-of-plane equations, so
that there is no loss of generality in taking $U_3 = 0$. The
incremental equations of motion and incompressibility
\eqref{motion} and \eqref{BC} then specialize to
\begin{equation} \label{motion1}
{\rm i}S_{11} + S'_{21} = - XU_1, \quad {\rm i}S_{12} + S'_{22} =
- XU_2, \quad {\rm i}U_1 + U'_2 = 0,
\end{equation}
where $X := \rho v^2$ and a prime signifies differentiation with
respect to the argument $k\hat{x}_2$.

It is convenient to introduce the quantities
\begin{equation} \label{tau}
\tau_1 = -{\rm i} S_{21}, \quad \tau_2 = -{\rm i} S_{22},
\end{equation}
which are functions of $\hat{x}_2$ proportional to the incremental
tractions on the planes parallel to the boundary.  Thus, from the
incremental boundary conditions \eqref{BC}, we have
\begin{equation} \label{BC1}
\tau_1(0) = \tau_2(0) = 0.
\end{equation}
The expressions
\begin{align} \label{Scomponents}
& S_{11} = {\rm i} (\hat{B}_{1111}+p)U_1  + {\rm i}\hat{B}_{1112}
U_2
           + \hat{B}_{1121}U_1' +  \hat{B}_{1122} U_2' - {\rm i}P,
\nonumber \\
& {\rm i}\tau_2= {\rm i} \hat{B}_{1122}U_1  + {\rm
i}\hat{B}_{1222} U_2
           + \hat{B}_{2122}U_1' +  (\hat{B}_{2222}+p) U_2' - {\rm i}P,
\nonumber \\
& {\rm i}\tau_1= {\rm i} \hat{B}_{1121}U_1  + {\rm
i}(\hat{B}_{1221}+p) U_2
           + \hat{\gamma}  U_1' +  \hat{B}_{2122} U_2',
\nonumber \\
& S_{12} = {\rm i} \hat{B}_{1112}U_1  + {\rm i}   \hat{\alpha} U_2
           + (\hat{B}_{1221}+p)U_1' +   \hat{B}_{1222} U_2'
\end{align}
may be derived from equations \eqref{wave}, \eqref{fields} and
\eqref{s}.

The short-hand notations defined by
\begin{align}\label{nus}
& \hat{\nu}_{12} = \hat{B}_{1222} - \hat{B}_{1112}
  = \textstyle{\frac{1}{2}}
     [\gamma - \alpha + (2\beta - \alpha - \gamma)\cos 2\theta]
        \sin 2 \theta,
\nonumber \\
& \hat{\nu}_{21} = \hat{B}_{1121} - \hat{B}_{2122}
  = \textstyle{\frac{1}{2}}
     [\gamma - \alpha - (2\beta - \alpha - \gamma)\cos 2\theta]
        \sin 2 \theta,
\end{align}
prove useful in rewriting the equations of motion in a convenient
form.  Indeed, a few manipulations and use of \eqref{motion1},
\eqref{Scomponents} and \eqref{nus} lead to the first-order
differential system
\begin{equation} \label{1stOrder}
 {\bf\mbox{\boldmath $\xi$}'}
    ={\rm i} \mathbf{N} \mbox{\boldmath $\xi$},
\quad
\mbox{\boldmath $\xi$}:=
   [U_1, U_2, \tau_1, \tau_2]^T,
\end{equation}
where
\begin{align} \label{N}
& \mathbf{N} = \begin{bmatrix}
                  \mathbf{N}_1 & \mathbf{N}_2 \\
      \mathbf{K}^{(1)} & \mathbf{N}_1^T
               \end{bmatrix},
 \quad  \mathbf{N}_1
     = \begin{bmatrix}
 -\hat{\nu}_{21}/\hat{\gamma}
        & - (\hat{\gamma} - \hat{\sigma}_{22})/\hat{\gamma}\\
                           - 1 & 0
       \end{bmatrix},
 \quad \mathbf{N}_2
   = \begin{bmatrix}
       1/\hat{\gamma} & 0  \\
                         0 & 0
     \end{bmatrix},
     \nonumber\\
 & \mathbf{K}^{(1)} = \begin{bmatrix}
    X - 2(\hat{\beta} + \hat{\gamma} - \hat{\sigma}_{22})
          +  \hat{\nu}^2_{21}/\hat{\gamma}
    &  \hat{\nu}_{12} +
  \hat{\nu}_{21} (\hat{\gamma}-\hat{\sigma}_{22})/\hat{\gamma}
 \\
    \hat{\nu}_{12} +
  \hat{\nu}_{21} (\hat{\gamma}-\hat{\sigma}_{22})/\hat{\gamma}
        & X  - \hat{\alpha}
           + (\hat{\gamma}-\hat{\sigma}_{22})^2/\hat{\gamma}
                                     \end{bmatrix}.
\end{align}

The classical approach to solving the equations of motion
\eqref{1stOrder} subject to \eqref{BC1} consists of writing the
solution as a decaying exponential function, such as
\begin{equation}
\mbox{\boldmath $\xi$}(kx_2) = \mbox{\boldmath $\xi$}^0e^{{\rm i}q
kx_2}, \quad \Im (q)>0,
\end{equation}
and then solving the eigenvalue problem $\mathbf{N}\mbox{\boldmath
$\xi$}^0 = q \mbox{\boldmath $\xi$}^0$. Its solution is of the
form
\begin{equation}
\mbox{\boldmath $\xi$}(kx_2) =
  A_1\mbox{\boldmath $\xi$}^1e^{{\rm i}q_1 kx_2}
   + A_2\mbox{\boldmath $\xi$}^2e^{{\rm i}q_2 kx_2},
\end{equation}
where $A_1$, $A_2$ are constants such that \eqref{BC1} is
satisfied. Explicitly, $q_1$ and $q_2$ are the complex roots of
the propagation condition $\det (\mathbf{N} - q\mathbf{I}) = 0$
with positive imaginary parts, where \textbf{I} is the
fourth-order identity.  The resulting equation is the quartic
\begin{equation} \label{propCond}
\hat{\gamma} q^4 + 2\hat{\nu}_{21}q^3 + (2\hat{\beta} - X) q^2
  + 2\hat{\nu}_{12}q + \hat{\alpha} - X =0
\end{equation}
for $q$, while $\mbox{\boldmath $\xi$}^i$, $i=1,2$ may be taken as
proportional to any column vector of the matrix adjoint to
$\mathbf{N} - q_i \mathbf{I}$.  If we take the third such column,
for example, then
\begin{equation}
 \mbox{\boldmath $\xi$}^i =
[q_i^2, -q_i,
  q_i(\hat{\gamma}q_i^2 + \hat{\nu}_{21}q_i
                 - \hat{\gamma} + \hat{\sigma}_{22}),
  -(\hat{\gamma}-\hat{\sigma}_{22})q_i^2 + \hat{\nu}_{12}q_i
                 + \hat{\alpha} - X]^T.
\end{equation}
The boundary conditions $\mbox{\boldmath $\xi$}(0) = [U_1(0),
U_2(0), 0, 0]^T$ then yield a homogeneous system of two linear
equations for the two unknowns $A_1$, $A_2$, whose determinant
must be zero.  This leads to the \textit{secular equation for the
surface wave speed}, written in the implicit form
\begin{multline} \label{seculImplicit}
  \hat{\gamma}(\hat{\alpha}-X)(q_1^2 + q_1q_2 + q_2^2)
 - (\hat{\gamma} - \hat{\sigma}_{22})
               (\hat{\gamma}q_1^2 q_2^2 + \hat{\alpha}-X) \\
+  [\hat{\nu}_{12}\hat{\nu}_{21}
      - (\hat{\gamma}-\hat{\sigma}_{22})^2] q_1q_2
 + [\hat{\nu}_{12}\hat{\gamma}q_1 q_2 + \hat{\nu}_{21}(\hat{\alpha}-X)](q_1 + q_2)
  = 0.
\end{multline}
It remains implicit as long as $q_1$ and $q_2$ are not expressed
explicitly.

%
\section{Explicit secular equations}
%
Here the problem of an incremental surface wave propagating over the
surface of a stretched and sheared incompressible half-space is
solved, up to the explicit derivation of the secular equation giving
the speed of the Rayleigh wave.

\subsection{Secular equation for materials with
$2\beta = \alpha + \gamma$}

Consider the Mooney-Rivlin strain-energy function
\begin{equation} \label{MR}
 W = \textstyle{\frac{1}{2}}
        C(\lambda_1^2 + \lambda_2^2 + \lambda_3^2 - 3)
       + \textstyle{\frac{1}{2}}
          D(\lambda_1^2\lambda_2^2 + \lambda_2^2\lambda_3^2
                                 + \lambda_3^2\lambda_1^2 - 3),
\end{equation}
where $C$ and $D$ are constant material parameters. For
\eqref{MR}, $\alpha$, $\gamma$ and $\beta$ reduce to
\begin{equation} \label{specialClass}
\alpha = \lambda_1^2 (C + D\lambda_3^2), \quad
\gamma = \lambda_2^2 (C + D\lambda_3^2), \quad
2\beta = (\lambda_1^2 + \lambda_2^2) (C + D\lambda_3^2),
\end{equation}
and so it is apparent that Mooney-Rivlin materials belong to the
general class of materials with energy function such that
\begin{equation} \label{classMR}
 2\beta = \alpha + \gamma.
\end{equation}
This class has the remarkable property that the propagation
condition for inhomogeneous plane waves with displacement
proportional to $\Re \{ \exp [{\rm i}k(\hat{x}_1 + q \hat{x}_2
-vt)] \}$ always admits $q^2+1$ as a factor when the body is
maintained in an arbitrary state of static homogeneous
deformation. In fact, Pichugin \cite{Pich03} proved that this
class is the most general one admitting this factorization.

Returning to the propagation condition \eqref{propCond} and using
\eqref{theta}, we find that when \eqref{classMR} holds the
quantities appearing in the quartic \eqref{propCond} simplify to
\begin{align} \label{hatMR}
  & \hat{\alpha}
  = \gamma \lambda_2^{-2}\left(\mu_1^2 + \mu_2^2 \kappa^2\right),
\quad
 \hat{\gamma} = \gamma \lambda_2^{-2} \mu_2^2,
\quad
 \hat{\nu}_{21} = \hat{\nu}_{12}
                     = -\gamma \lambda_2^{-2} \mu_2^2 \kappa,
\nonumber \\
 & 2 \hat{\beta}
     = \gamma \lambda_2^{-2}
           \left[\mu_1^2 + \mu_2^2 \left(1+\kappa^2\right)\right],
\quad X = \gamma \lambda_2^{-2}\mu_2^2\left(\mu_1^2/\mu_2^2 -
\eta^2\right),
\end{align}
where the quantity
\begin{equation}
  \eta :=  \dfrac{\mu_1}{\mu_2}
   \sqrt{1- \dfrac{\lambda_2^2}{\mu_1^2}\dfrac{X}{\gamma}}
\end{equation}
has been introduced in the last equality. Then, after removal of
the factor $\hat{\gamma}\neq 0$, the quartic \eqref{propCond}
factorizes as expected to give
\begin{equation} \label{factorize}
(q^2+1) (q^2 - 2\kappa q + \kappa^2 + \eta^2) =0,
\end{equation}
and its roots with positive imaginary parts are
\begin{equation} \label{q1q2MR}
  q_1={\rm i}, \quad q_2= \kappa + {\rm i} \eta.
\end{equation}
In passing, we note that the condition $\Im (q_i)>0$ for decay of
the wave yields an upper bound on the admissible wave speed, such
that
\begin{equation}
  0 \le X = \rho v^2 \le \mu_1^2 \lambda_2^{-2} \gamma,
\end{equation}
with the limiting value $\mu_1^2 \lambda_2^{-2}\gamma$
corresponding to a body wave not a surface wave (which requires
strict inequality).

Finally, substituting the values \eqref{q1q2MR} for $q_1$ and
$q_2$ into \eqref{seculImplicit}, we arrive at the
\textit{explicit secular equation}
\begin{equation} \label{secularMR}
  \eta^3 + \eta^2 +(3+\kappa^2 - 2\bar{\sigma}_{22})\eta
    - (1-\bar{\sigma}_{22})^2 =0,
\end{equation}
where $\bar{\sigma}_{22}
   := \lambda_2^2 \hat{\sigma}_{22} / (\mu_2^2 \gamma)$.
This secular equation is identical (apart from notational
differences) to that obtained when the material is stretched but
not sheared ($\kappa=0$) \cite{Flav63,DoOg90} or when the material
is sheared but not stretched ($\mu_i=1$) \cite{CoOg95}.

In the case of the Mooney-Rivlin strain energy \eqref{MR}, $\gamma
\lambda_2^{-2} = C + D/(\mu_1 \mu_2)^2$, so that both $\eta$ and
$\bar{\sigma}_{22} $ are independent of the amount of shear
$\kappa$.
Of course $\eta_R$, the positive real root of the cubic
\eqref{secularMR}, depends on $\kappa$.
By differentiation we find that
$\partial \eta_R / \partial (\kappa^2) =
 -\eta_R^2 / [2\eta_R^3 + \eta_R^2 + (1-\bar{\sigma}_{22})^2]<0$ and
so that, by \eqref{hatMR}$_4$, the surface wave speed is a
monotone increasing function of the amount of shear. Hence for
Mooney-Rivlin materials subject to a given triaxial stretch
$\mu_1, \mu_2, \mu_3$, the greater the subsequent shear is, the
faster the Rayleigh wave propagates.

In the limit $X=0$ the secular equation  \eqref{secularMR} reduces to
the \textit{bifurcation equation}.  This defines a curve in ($\mu_i,
\kappa, \bar{\sigma}_{22}$)-space that separates the region where
the surface wave exists (and is unique) and the half-space is
stable from the region where the surface wave does not exist and
the half-space might become unstable (see Chadwick and Jarvis
\cite{ChJa79} or Dowaikh and Ogden \cite{DoOg90}).  The
bifurcation equation is
\begin{equation} \label{bifurcationMR}
  \left(\dfrac{\mu_1}{\mu_2}\right)^3
        + \left(\dfrac{\mu_1}{\mu_2}\right)^2
   +(3+\kappa^2 - 2\bar{\sigma}_{22})\left(\dfrac{\mu_1}{\mu_2}\right)
    - (1-\bar{\sigma}_{22})^2 =0.
\end{equation}
When there is no normal Cauchy pre-stress in the $\hat{x}_2$
direction ($\bar{\sigma}_{22}=0$), the bifurcation criterion
becomes universal with respect to the entire class of
incompressible materials satisfying \eqref{classMR} since it does
not depend on the material parameters associated with a specific
strain energy, but only on the pre-stretch and the amount of
shear.

\subsection{Secular equation for a general strain energy}

For a general (incompressible, isotropic) strain-energy function
($2\beta \ne \alpha + \gamma$), the propagation condition
\eqref{propCond} does not factorize as in \eqref{factorize} and it
is difficult in general to use the quartic \eqref{propCond} for the
derivation of an explicit secular equation.  Instead, we employ
the method of the `polarization vector' devised by Currie
\cite{Curr79} and based on the fundamental equations
\begin{equation} \label{fundamental}
  \overline{\mathbf{U}}(0) \cdot [\mathbf{K}^{(n)} \mathbf{U}(0)]=0.
\end{equation}
Here, the symmetric $2 \times 2$ matrix $\mathbf{K}^{(n)}$ is the
lower left block of $\mathbf{N}^n$, where $n$ is any positive or
negative integer. Thus, for instance, $\mathbf{K}^{(1)}$ is given
by \eqref{N}$_4$, and we find that $\mathbf{K}^{(-1)}$ and
$\mathbf{K}^{(2)}$ are given by
\begin{equation}
 \frac{1}{X-\hat{\alpha}}
  \begin{bmatrix}
\hat{\gamma}(X-\hat{\alpha}) + (\hat{\gamma}-\hat{\sigma}_{22})^2
  &
-\hat{\nu}_{21}(X-\hat{\alpha})
    + \hat{\nu}_{12}(\hat{\gamma}-\hat{\sigma}_{22}) \\
 -\hat{\nu}_{21}(X-\hat{\alpha})
    + \hat{\nu}_{12}(\hat{\gamma}-\hat{\sigma}_{22})  &
\hat{\nu}_{12}^2 - (X-\hat{\alpha})
    (X-2\hat{\beta}-2\hat{\gamma}+2\hat{\sigma}_{22})
  \end{bmatrix},
 \end{equation}
and
\begin{equation}
- \begin{bmatrix} 2\hat{\nu}_{21}K^{(1)}_{11}/\hat{\gamma} +
2K^{(1)}_{12}
  &
\hat{\nu}_{21}K^{(1)}_{12}/\hat{\gamma} + K^{(1)}_{22}
 +(\hat{\gamma}-\hat{\sigma}_{22})K^{(1)}_{11}/\hat{\gamma} \\
 \hat{\nu}_{21}K^{(1)}_{12}/\hat{\gamma} + K^{(1)}_{22}
 +(\hat{\gamma}-\hat{\sigma}_{22})K^{(1)}_{11}/\hat{\gamma}
 &
2(\hat{\gamma}-\hat{\sigma}_{22})K^{(1)}_{12}/\hat{\gamma}
  \end{bmatrix},
 \end{equation}
respectively. Here $K^{(2)}_{11}$ and $K^{(2)}_{12}$ are linear in
$X$ and $K^{(2)}_{22}$ is independent of $X$.

Now, the fundamental equations \eqref{fundamental}, written in
turn for $n=1,-1,2$, yield the homogeneous system
\begin{equation}\label{Kmat}
  \begin{bmatrix}
   K^{(1)}_{11}  & K^{(1)}_{12}  & K^{(1)}_{22}  \\
   K^{(-1)}_{11} & K^{(-1)}_{12} & K^{(-1)}_{22} \\
   K^{(2)}_{11}  & K^{(2)}_{12}  & K^{(2)}_{22}
 \end{bmatrix}
  \begin{bmatrix}
    U_1(0) \overline{U}_1(0) \\
    U_1(0) \overline{U}_2(0) +  \text{c.c.}\\
    U_2(0) \overline{U}_2(0)
  \end{bmatrix}
= \begin{bmatrix} 0 \\ 0 \\ 0 \end{bmatrix},
\end{equation}
where c.c. denotes the complex conjugate of the preceding term.
The determinant of the left-hand matrix in \eqref{Kmat} must be
zero, and the resulting equation is the \textit{explicit secular
equation for surface waves in a stretched and sheared
incompressible material}. It is a quartic in $X$, but since its
explicit expression is very lengthy we do not give it here,
although it has been obtained using Maple and Mathematica. We
note, however, that in the stress-free configuration
($\hat{\sigma}_{22} =0, \mu_i=1, \kappa=0$) it factorizes into the
product of a term linear in $X$ and the cubic of Rayleigh
\cite{Rayl85} for isotropic incompressible elastic solids, namely
\begin{equation}
  \zeta^3 - 8\zeta^2 + 24\zeta - 16 =0, \quad
 \zeta:= \rho v^2 / \check{\gamma},
\end{equation}
which has a unique real root $\zeta_R \approx 0.9126$. Here
$\check{\gamma}>0$ is the shear modulus of the incompressible
material in that configuration.

As an illustration, we consider the generalized Varga
strain-energy function
\begin{equation} \label{Varga}
 W = C(\lambda_1 + \lambda_2 + \lambda_3 - 3)
  + D(\lambda_1\lambda_2 + \lambda_2\lambda_3
       + \lambda_3\lambda_1 - 3),
\end{equation}
where again $C$ and $D$ are constant material parameters, but
different from those in \eqref{MR}.  For discussions of
\eqref{Varga} see, for example, Carroll \cite{Carr88},
Haughton \cite{Haug93}, or Hill \cite{Hill01}.
For \eqref{Varga}, $\alpha$, $\gamma$ and $\beta$ reduce to
\begin{equation} \label{specialVarga}
\alpha =
   \dfrac{\lambda_1^2 (C + D\lambda_3)}
      {\lambda_1+\lambda_2}, \quad
\gamma =
   \dfrac{\lambda_2^2 (C + D\lambda_3)}
      {\lambda_1+\lambda_2}, \quad
\beta =
   \dfrac{\lambda_1\lambda_2(C + D\lambda_3)}
      {\lambda_1+\lambda_2},
\end{equation}
and it is then apparent that \eqref{Varga} belongs to the class of
materials with energy function such that
\begin{equation} \label{classVarga}
  \beta = \sqrt{\alpha  \gamma}.
\end{equation}
For such materials we find first that, on use of \eqref{theta},
\begin{align}
 & \hat{\alpha} = \dfrac{\gamma}{\lambda_2^2(\lambda_1+\lambda_2)^2}
   (\mu_1^2 + \mu_1 \mu_2 + \mu_2^2 \kappa^2)^2,
\quad
 \hat{\gamma} = \dfrac{\gamma}{\lambda_2^2(\lambda_1+\lambda_2)^2}
   (\mu_1 \mu_2 + \mu_2^2)^2,
\nonumber \\
 &  \hat{\beta} = \dfrac{\gamma}{\lambda_2^2(\lambda_1+\lambda_2)^2}
   [\mu_1 \mu_2(\mu_1 + \mu_2)^2
                + \mu_2^2 \kappa^2(\mu_1 \mu_2 + 3\mu_2^2)],
\nonumber \\
 & \hat{\nu}_{12}
      =  -2 \dfrac{\gamma}{\lambda_2^2(\lambda_1+\lambda_2)^2}
   \mu_2^2 \kappa(\mu_1^2 + \mu_1 \mu_2 + \mu_2^2
   \kappa^2),\nonumber\\
 & \hat{\nu}_{21}
     = -2 \dfrac{\gamma}{\lambda_2^2(\lambda_1+\lambda_2)^2}
   \mu_2^2 \kappa(\mu_1 \mu_2 + \mu_2^2).
\end{align}

Secondly, we find that, \textit{in the absence of normal load\/}
($\hat{\sigma}_{22} = 0$) on the plane boundary in the underlying
configuration, the secular equation simplifies to
\begin{align} \label{secularVarga}
2(\mu_1 + \mu_2) x^4
 - (3\mu_1+7\mu_2)\mu^4 x^3
  + [(\mu_1+5\mu_2) \mu^2 + 2\mu_2(3\mu_1^2+5\mu_2^2)]\mu^6 x^2
\nonumber \\
   - 4 (\mu_1 + \mu_2)^2 \mu_2 \mu^{10} x
    + (3\mu_1 - \mu_2)(\mu_1+\mu_2)^2 \mu_2^2 \mu^{12} = 0,
\end{align}
where
\begin{equation}
x := \lambda_2^2(\lambda_1+\lambda_2)^2 X / \gamma,
\quad
\mu^2 := (\mu_1 + \mu_2)^2 + \mu_2^2\kappa^2.
\end{equation}
The \textit{bifurcation equation} becomes simply
\begin{equation} \label{bifurcationVarga}
3\mu_1 - \mu_2 = 0,
\end{equation}
which is universal for the class of materials satisfying
\eqref{specialVarga}, and is independent of the amount of shear,
i.e. it is the same bifurcation equation as for a material subject
only to a triaxial stretch \cite{DeSc04}. When the material is
stretched but not sheared ($\kappa=0$), the secular equation
factorizes into the product of a term linear in $x$ and the cubic
in the secular equation for principal surface waves \cite{DoOg90}.
When the material is sheared but not stretched ($\mu_i=1$), on the
other hand, it reduces to
 \begin{equation}
2 x^4 - 5\mu^4 x^3 + (8 + 3 \mu^2)\mu^6 x^2
  - 8 \mu^{10} x + 4 \mu^{12} = 0,
\quad \mu^2 = 4 + \kappa^2,
\end{equation}
with $x$ appropriately specialized.

Finally, we find that, \textit{in the presence of normal load\/}
($\hat{\sigma}_{22} \ne 0$) on the plane boundary in the
underlying configuration, the secular equation is too lengthy to
reproduce here, but the \textit{bifurcation equation} is
relatively simple and consists of the two possibilities
\begin{equation}
  \overline{\sigma}_{22}
    + \mu_2(3\mu_1-\mu_2)[(\mu_1+\mu_2)^2 + \mu_2^2\kappa^2] =
    0,\quad \overline{\sigma}_{22}=\mu^2\mu_2(\mu_1+\mu_2),
\end{equation}
where $\overline{\sigma}_{22}
 = \lambda_2^2 (\lambda_1+\lambda_2)^2 \hat{\sigma}_{22} / \gamma$.
For stability $\overline{\sigma}_{22}$ must lie between the values
specified by these two equations.  This result generalizes that in
Dowaikh and Ogden \cite{DoOg90} (their equation (7.15)), which is
applicable for $\kappa=0$, and recovers their result in this case.

%
\section{Numerical results}
%

\subsection{Mooney-Rivlin materials}

For materials such as the neo-Hookean or Mooney-Rivlin models
satisfying $2\beta = \alpha + \gamma$, the secular equation
\eqref{secularMR} for stretched and sheared bodies has been
written in a form which is formally identical to that obtained by
Connor and Ogden \cite{CoOg96} for sheared (but not stretched)
bodies, with the difference that their $\eta$ and
$\overline{\sigma}_{22}$ are a specialization of ours to the case
$\mu_1=\mu_2=\mu_3=1$. As a result the ($\eta,
\overline{\sigma}_{22}$) dispersion curves are common to those in
\cite{CoOg96} and there is little interest in reproducing them
here.

Instead we concentrate on the bifurcation condition
\eqref{bifurcationMR}. For a Mooney-Rivlin material subject to a
pure homogeneous deformation ($\kappa=0$) and no normal load
($\hat{\sigma}_{22} = 0$), Biot \cite{Biot65} found that
(linearized) surface instability occurs at a critical stretch of
$(\mu_1)_\text{cr} = 0.544$ for the plane strain $\mu_3=1$ and of
$(\mu_1)_\text{cr} = 0.444$ for the equibiaxial pre-strain
$\mu_2=\mu_3$, and Green and Zerna \cite{GrZe92} found a critical
stretch of $(\mu_1)_\text{cr} = 0.666$ for the equibiaxial
pre-strain $\mu_1=\mu_3$. For a stretched ($\mu_i \ne 1$) and
sheared ($\kappa \ne 0$) Mooney-Rivlin material subject to a
pre-load $\hat{\sigma}_{22} \ne 0$ we find from
\eqref{bifurcationMR} that $\mu_1 / \mu_2$ is a monotonic
decreasing function of the amount of shear $\kappa$. Hence, in the
cases where $\mu_1 / \mu_2$ is a monotonic increasing function of
$\mu_1$, we conclude that if the half-space is stable in the
stretched configuration, then it will remain stable in the
stretched and sheared configuration whatever the amounts of
pre-load and of shear are applied. These cases include the plane
strain $\mu_3=1$ (because then $\mu_1 / \mu_2 = \mu_1^2$) and the
equibiaxial pre-strains $\mu_2=\mu_3$, $\mu_1=\mu_3$ (because then
$\mu_1 / \mu_2 = \mu_1^{3/2}, \mu_1^3$, respectively). Figures
2(a) and 2(b) show indeed this behaviour for $(\mu_1)_\text{cr}$
in the plane strain case, for different values of the scaled
normal load $\overline{\sigma}_{22}$.

\begin{figure}[!t]
\centering
\mbox{\subfigure{\epsfig{figure=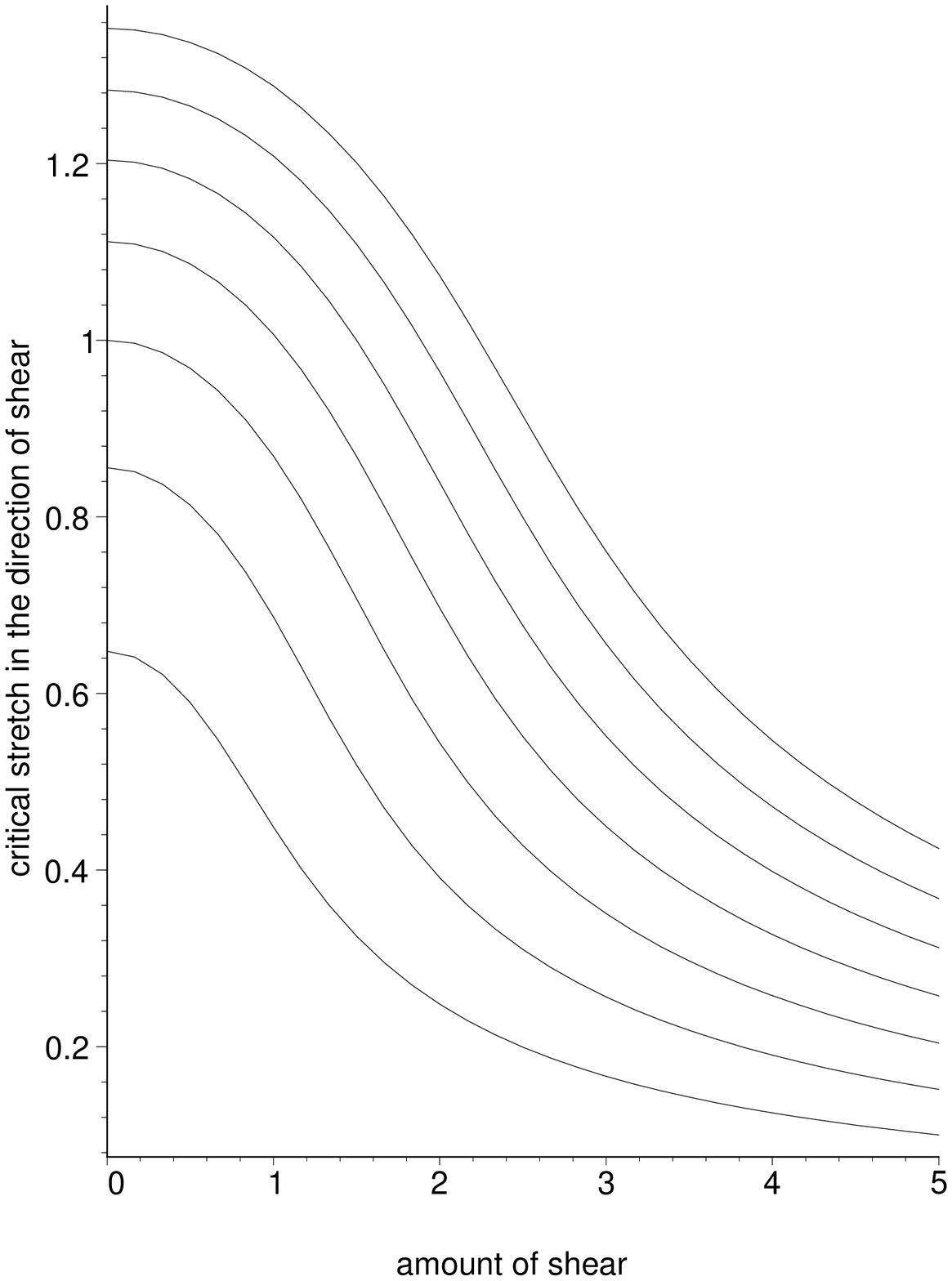,
width=.5\textwidth}}
  \quad \quad
     \subfigure{\epsfig{figure=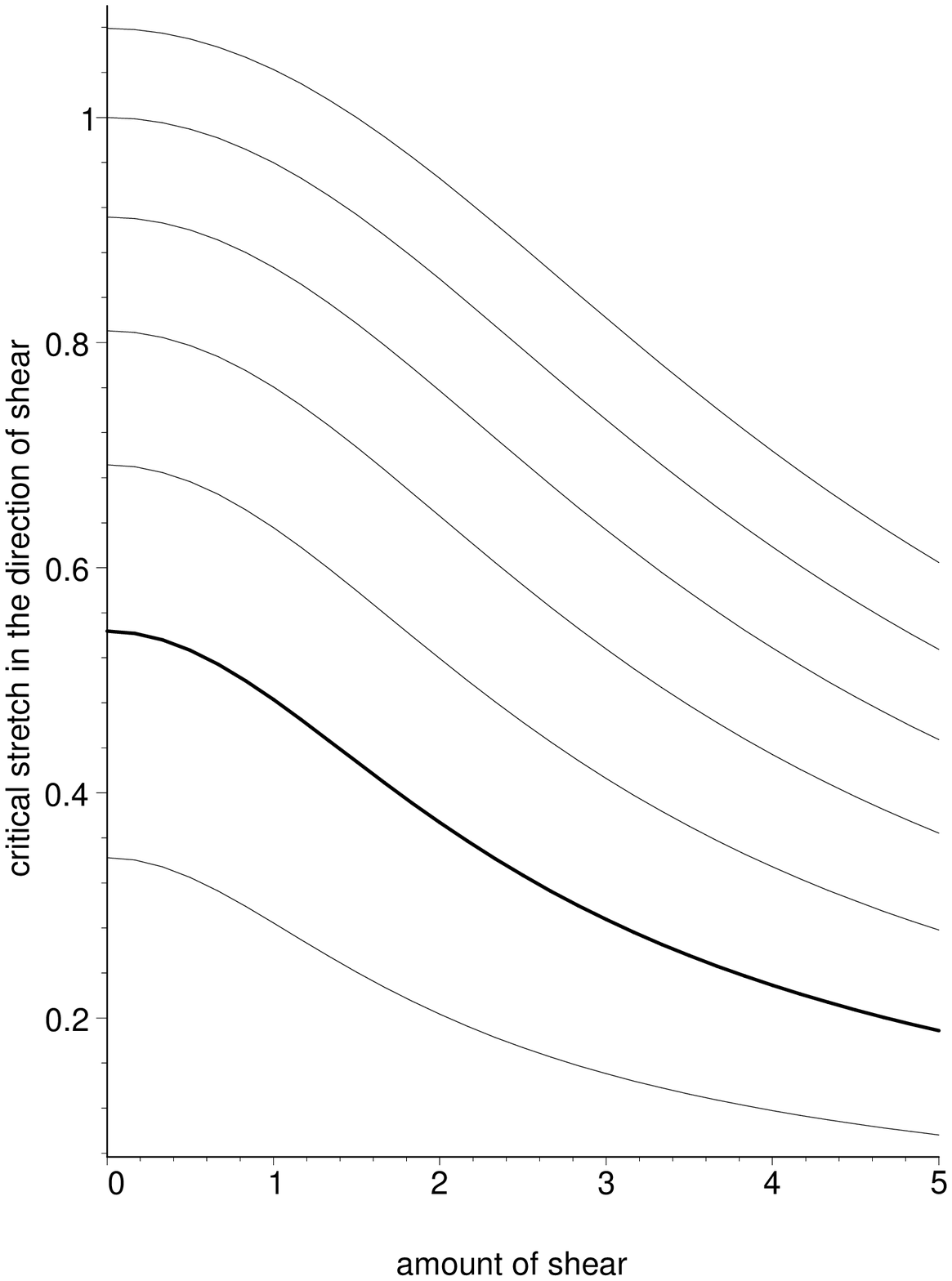,
width=.5\textwidth}}} \caption{Plot of $(\mu_1)_\text{cr}$ as a
function of $\kappa$ for a Mooney-Rivlin material in plane strain
under the normal loads (from top curve  to bottom curve) (a)
$\overline{\sigma}_{22} = 1.5, 1.75, 2, 2.25, 2.5, 2.75, 3$; and
(b) $\overline{\sigma}_{22} =  -2.5, -2, -1.5, - 1, - 0.5, 0$
(thick curve) $0.5$.}
\end{figure}

%

\subsection{Varga materials}

Now we turn our attention to a half-space made of generalized
Varga material \eqref{Varga}, subject to large stretches and shear
such that no extension occurs in the direction normal to the plane
of shear. Hence $\mu_3 = \lambda_3 = 1$ and $\mu_2 = \mu_1^{-1}$.
When there is no normal load ($\hat{\sigma}_{22} = 0$), then the
critical stretch found from \eqref{bifurcationVarga} is
$(\mu_1)_\text{cr} = 0.577$ \cite{DeSc04}. As $\mu_1$ is increased
we find that the wave speed first increases and then decreases,
whatever the amount of superposed shear, with $c_s :=
\sqrt{(C+D)/(2\rho)}$, the speed of shear waves in the undeformed
material, as an asymptotic limit. In contrast to Mooney-Rivlin
materials, the presence of shear decreases the surface wave speed
in Varga materials subject to a given (fixed) amount of triaxial
stretch $\mu_1, \mu_2, \mu_3$. Figure 3(a) illustrates these
features, with a plot of the squared surface wave speed, scaled
with respect to $c_s^2$, and traced as a function of $\mu_1$ for
different values of $\kappa$.
At $\kappa=0$, $\mu_1=1$, the material is unstressed and we recover
$v^2 / c_s^2 = 0.9126$, as indicated by the dotted lines.
We record here that for some values
of the parameters equation \eqref{secularVarga} may have two
solutions for $x$.  However, because of the rationalization
implicit in the derivation of \eqref{secularVarga}, the second
solution is spurious and is not associated with a surface wave.

Next we exploit the remarkable property of the deformation
\eqref{static} that it allows a square block made of a
hyperelastic material to be stretched and sheared without either
shear tractions or \textit{normal} tractions applied on its faces
\cite{RaWi87}. This possibility arises when $\mu_1^2 =
(1+\kappa^2)\mu_2^2$. The actual dependence of the $\mu_i$ on
$\kappa$ is then found from the constitutive equation of a given
material. For the neo-Hookean model, Rajagopal and Wineman
\cite{RaWi87} found that
\begin{equation} \label{muBlock}
  \mu_1 = (1+\kappa^2)^{\frac{1}{3}}, \quad
  \mu_2 = \mu_3 = (1+\kappa^2)^{-\frac{1}{6}}.
\end{equation}
It can be checked that the same condition applies for
Mooney-Rivlin and for generalized Varga materials. We computed the
speed of a surface wave travelling over such a block, made of a
classical Varga material ($D=0$ in \eqref{Varga}). Figure 3(b)
displays the variation of the squared wave speed, scaled with
respect to the squared shear wave speed $c_s^2 = C/(2\rho)$ in the
undeformed material, as a function of the amount of shear. As
$\kappa$ increases, the $\mu_i$ vary according to \eqref{muBlock},
and the wave speed increases in a monotone manner, at least within
the `reasonable' range $[0,5]$ for $\kappa$. For greater, but
unrealistic, values of $\kappa$ we found that the wave speed $v$
eventually decreases, with $c_s$ as an asymptotic limit.

\begin{figure}[!t]
\centering
\mbox{\subfigure{\epsfig{figure=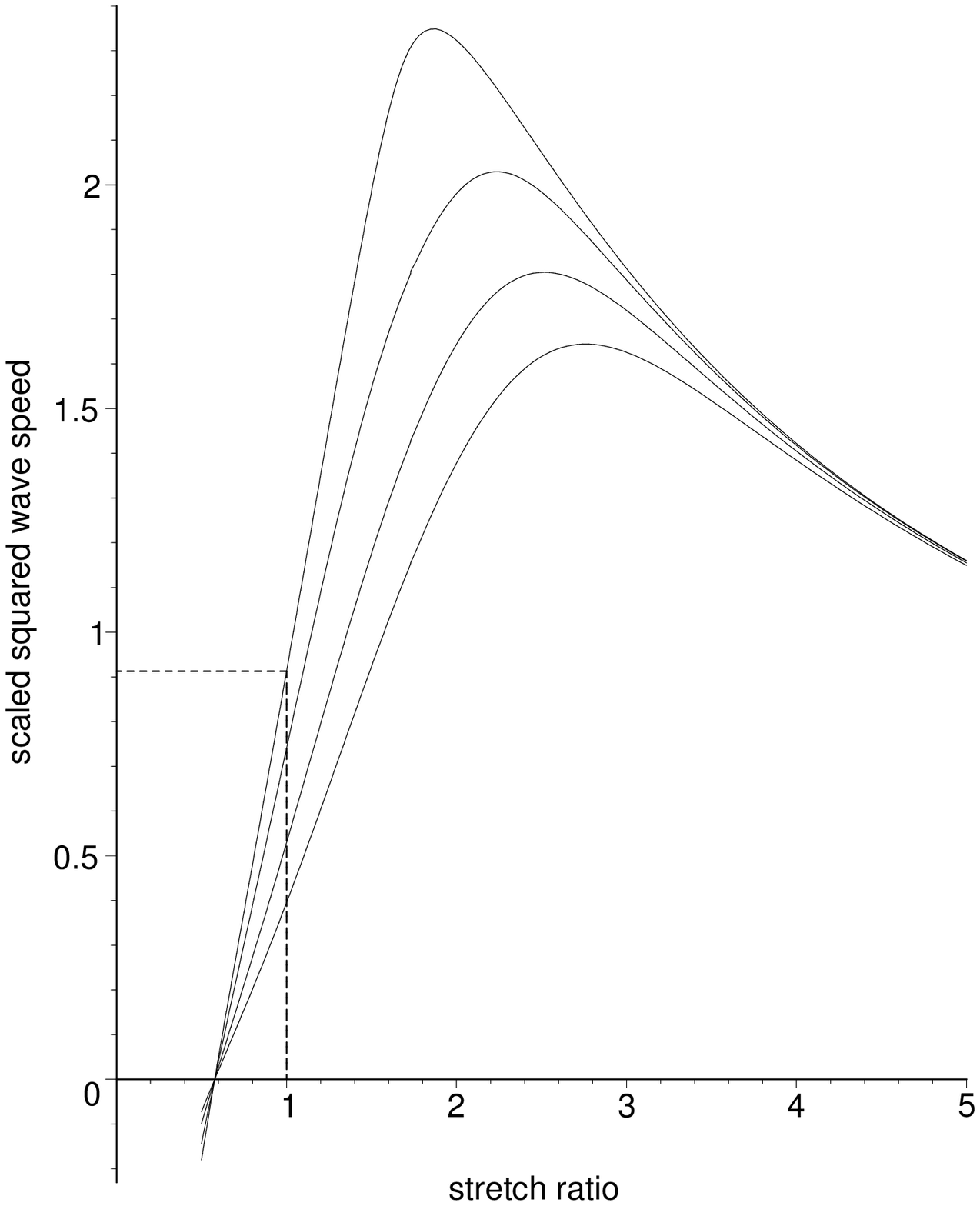,
width=.5\textwidth}}
  \quad \quad
     \subfigure{\epsfig{figure=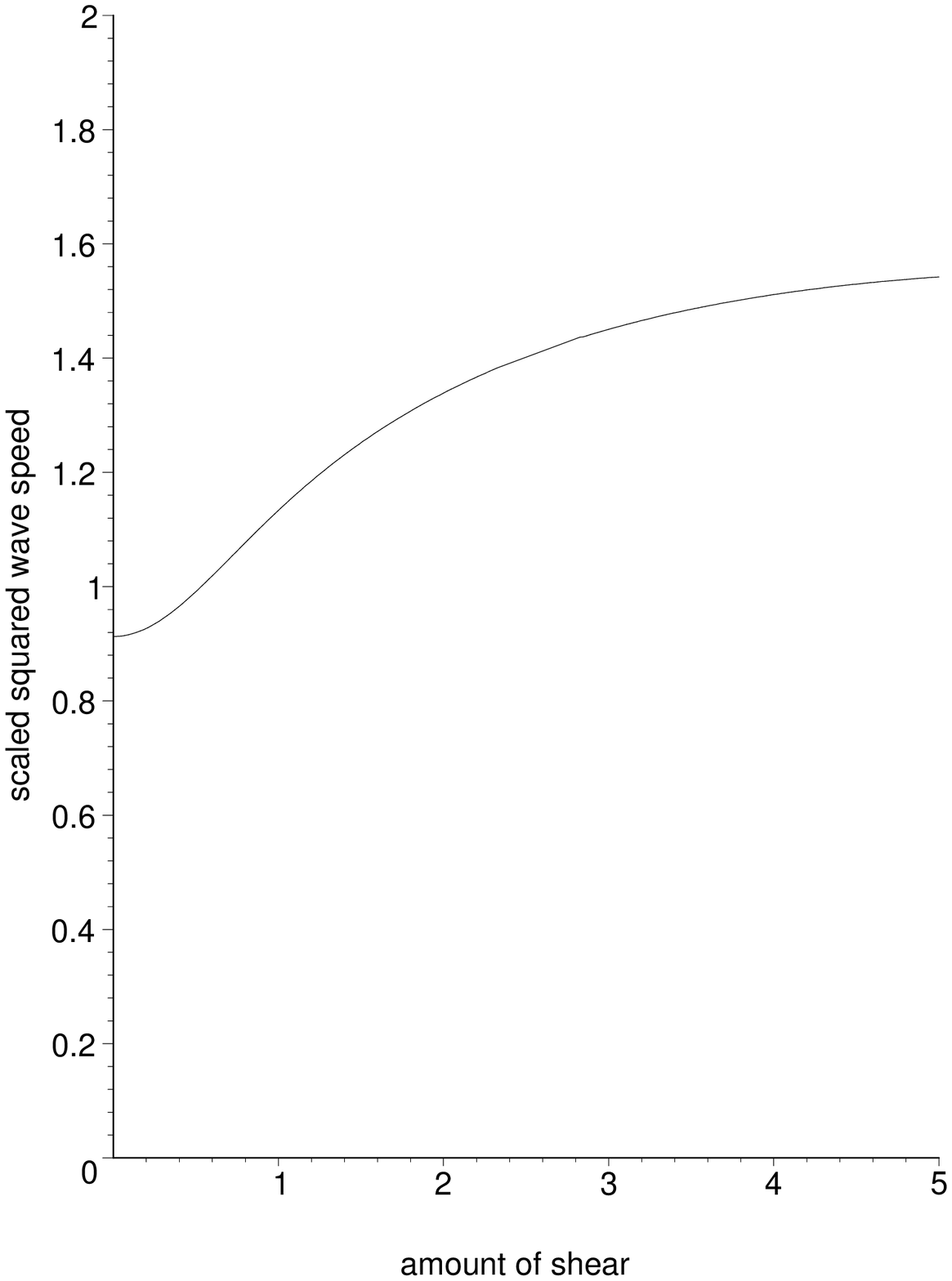,
width=.5\textwidth}}} \caption{Plot of $v^2/c_s^2$ (a) against
$\mu_1$ for (from top curve  to bottom curve) $\kappa = 0,1,2,3$,
and (b) against $\kappa$ in plane strain for a square block of
Varga material.}
\end{figure}

\subsection{Sheared Gent materials}

We conclude with the study of the strain-energy function proposed
by Gent \cite{Gent96}, namely
\begin{equation} \label{Gent}
W = \textstyle{\frac{1}{2}}
  C J_m
   \ln \left(1 -
    \dfrac{\lambda_1^2 + \lambda_2^2 + \lambda_3^2 - 3}{J_m} \right),
\quad \lambda_1^2 + \lambda_2^2 + \lambda_3^2 < 3 + J_m,
\end{equation}
that accounts for the limiting chain extensibility of some elastic
materials such as rubber or soft biological tissue. Here $C$ is
the infinitesimal shear modulus and $J_m$ is a constant. Recently,
Horgan and Saccomandi \cite{HoSa03} showed how and to what extent
this model could be used to describe finite deformations of
strain-stiffening biological tissues, such as aortas and arteries.
Typically, these biomaterials can be subjected to quite large
deformations (within certain limitations) and exhibit highly
nonlinear constitutive equations. These features are quite well
accounted for by the strain-energy function \eqref{Gent}: it is
nonlinear in the stretches, the limiting condition
\eqref{Gent}$_2$ imposes an upper bound on the value of the
stretch, and the two parameters $C$ and $J_m$ can be adjusted to
render a satisfactory picture of many deformations and of the
strain-hardening process (see \cite{HoSa03} and the references
therein to other works by Horgan and collaborators for extensive
discussion of and justification for the use of the Gent strain
energy-function for the modelling of stress-hardening biological
tissues). Using classical experimental data \cite{LaKi51} on the
aorta of a 21 year old male and on the aorta of a 70 year old
male, Horgan and Saccomandi computed values of $J_m = 2.289,
0.422$ for the younger and older aorta, respectively.

Here we focus on a body composed of Gent material subject to a
finite shear ($\mu_i =1$, $\kappa >0$ in \eqref{static}). We see
at once from (2.7) that the limiting chain extensibility condition
\eqref{Gent}$_2$ yields an upper bond for the amount of shear and
thus for the angle of shear,
\begin{equation}
\kappa^2 < J_m, \quad \phi < \tan^{-1} (\sqrt{J_m}).
\end{equation}
Hence, using the values of $J_m$ given above, we find that the
younger aorta cannot be sheared beyond an amount of shear of
approximately 1.513, corresponding to an angle of shear of about
56.54$^\circ$. For the older (stiffer) aorta, the corresponding
values are: 0.650 and 33.01$^\circ$, respectively.

Now turning back to small amplitude motions superposed on a large
homogeneous deformation with principal stretches $\lambda_1,
\lambda_2, \lambda_3$, we find that for Gent materials the elastic
moduli (2.14) are given by
\begin{align}
& \alpha = C \dfrac{J_m}
  {J_m + 3 - \lambda_1^2 - \lambda_2^2 - \lambda_3^2}
     \lambda_1^2, \quad
  \gamma = C \dfrac{J_m}
  {J_m + 3 - \lambda_1^2 - \lambda_2^2 - \lambda_3^2}
     \lambda_2^2,
\nonumber \\
& 2 \beta = C \dfrac{J_m}
  {J_m + 3 - \lambda_1^2 - \lambda_2^2 - \lambda_3^2}
     \left[ \lambda_1^2 + \lambda_2^2
 +  \dfrac{2(\lambda_1^2 - \lambda_2^2)^2}
  {J_m + 3 - \lambda_1^2 - \lambda_2^2 - \lambda_3^2}
\right].
\end{align}
When the homogeneous deformation is a shear with amount $\kappa$,
they reduce to
\begin{align}
& \alpha = \dfrac{C}{4} \dfrac{J_m}{J_m - \kappa^2}
    (\sqrt{4+\kappa^2} + \kappa)^2,  \quad
  \gamma = \dfrac{C}{4} \dfrac{J_m}{J_m - \kappa^2}
    (\sqrt{4+\kappa^2} - \kappa)^2,
\nonumber \\
& 2 \beta = C \dfrac{J_m}{J_m - \kappa^2}
     \left[ 2 + \kappa^2
              +  \dfrac{2\kappa^2(4+ \kappa^2)}
                                {J_m - \kappa^2}\right].
\end{align}
These expressions, together with the relations \eqref{theta},
evaluated for $\mu_1 = \mu_2 =1$, allow for the explicit
computation of the quantities $\hat{\alpha}$,  $\hat{\gamma}$,
$\hat{\beta}$ appearing in the secular equation. By
\eqref{shortHand}, we have
\begin{align}
& \hat{\alpha} = C \dfrac{J_m}{J_m - \kappa^2}
    \left[1 + \kappa^2   + \dfrac{2\kappa^2}{J_m - \kappa^2}\right],
&&
  2 \hat{\beta} = C \dfrac{J_m}{J_m - \kappa^2}
    \left[2 + \kappa^2   - \dfrac{2\kappa^2(2 - \kappa^2)}
                       {J_m - \kappa^2}\right],
\nonumber \\
& \hat{\gamma} = C \dfrac{J_m}{J_m - \kappa^2}
    \left[1 + \dfrac{2\kappa^2}{J_m - \kappa^2}\right],
&&
   \hat{\nu}_{12},\, \hat{\nu}_{21}
= -C \dfrac{J_m}{J_m - \kappa^2}
    \kappa \left[1 \mp  \dfrac{2\kappa^2}{J_m - \kappa^2}\right].
\end{align}
Now $\sqrt{\rho v^2/C}$, the surface wave speed scaled with
respect to the speed of a (bulk) shear wave propagating in the
undeformed material, can be plotted as a function of the amount of
shear $\kappa \in \left[0, \sqrt{J_m} \,\right)$. Figure 3 shows
that the surface wave speed increases monotonically with the
amount of shear, mildly for small shears and then dramatically as
the limiting amount of shear is approached (vertical asymptotes at
$\kappa = 1.513, 0.650$ for the younger and older aorta,
respectively). The curves are plotted in the absence of normal
load ($\hat{\sigma}_{22}=0$); for a discussion on the influence of
normal load on the wave speed and on the surface stability of
sheared incompressible materials, see Connor and Ogden
\cite{CoOg95}.

\begin{figure}[!t]
\centering \epsfig{figure=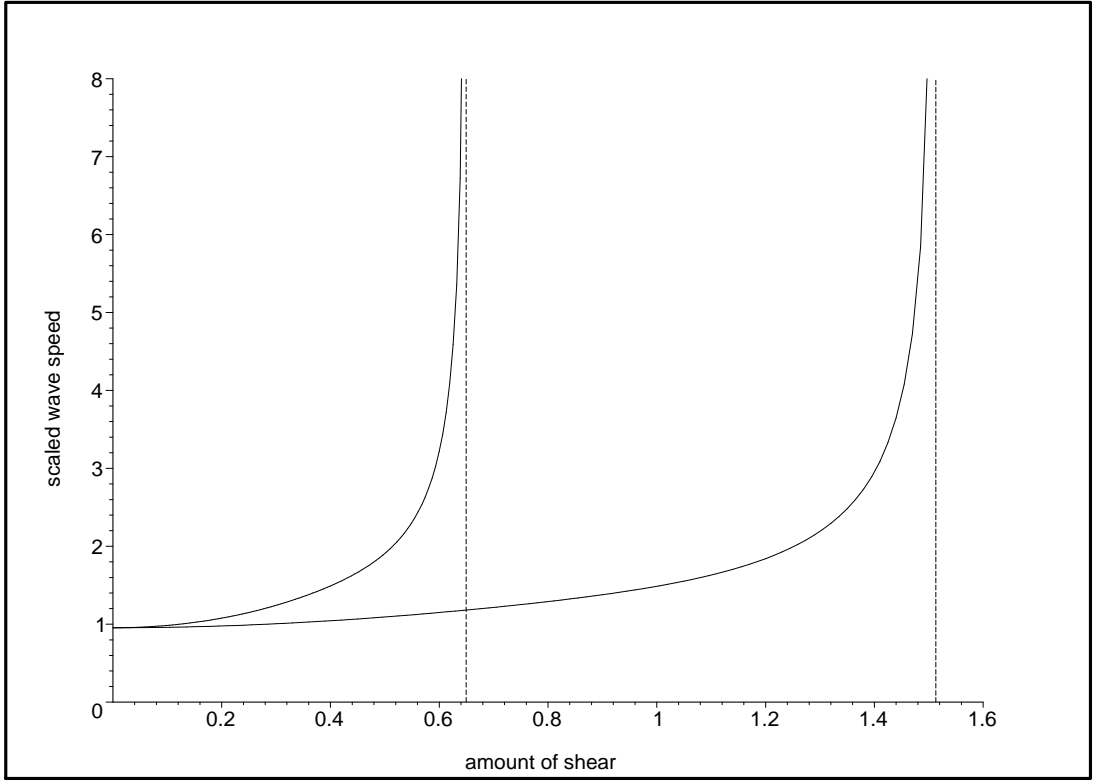,width=.7\textwidth,angle=-90}
\caption{Plot of the surface wave speed $\sqrt{\rho v^2/C}$ as a
function of the amount of shear $\kappa \in \left[0, \sqrt{J_m}
\,\right)$ for a Gent material: $\sqrt{J_m}=0.650$ (left curve);
$\sqrt{J_m}=1.513$ (right curve).}
\end{figure}


\end{document}